\documentclass[aps,prd,showpacs,nofootinbib,twocolumn]{revtex4-1}
\usepackage{latexsym}
\usepackage{amsmath,amsfonts}
\usepackage{amsbsy}
\usepackage{mathrsfs}
\usepackage{verbatim}
\usepackage{graphicx,calc,epsfig}
\def\ut#1{\rlap{\lower1ex\hbox{$\sim$}}#1{}}

\newcommand{\be}{\nopagebreak[3]\begin{equation}}
\newcommand{\ee}{\end{equation}}
\newcommand{\bee}{\nopagebreak[3]\begin{equation*}}
\newcommand{\eee}{\end{equation*}}
\newcommand{\ba}{\nopagebreak[3]\begin{eqnarray}}
\newcommand{\ea}{\end{eqnarray}}
\newcommand{\baa}{\nopagebreak[3]\begin{eqnarray*}}
\newcommand{\eaa}{\end{eqnarray*}}
\newcommand{\la}{\label}
\DeclareFontFamily{U}{rsfs}{}         
\DeclareFontShape{U}{rsfs}{m}{n}{<5> rsfs5 <6><7> rsfs7          %
  <8><9><10><10.95><12><14.4><17.28><20.74><24.88> rsfs10}{}     %
\DeclareMathAlphabet{\mathfs}{U}{rsfs}{m}{n}                     %
\newcommand{\mfs}[1]{\mathfs {#1}}                               %
\newcommand{\va}{\scriptscriptstyle}

\newcommand{\n}{{\nonumber}}

\newcommand{\sH}{{\mfs H}}
\newcommand{\sL}{{\mfs L}}

\newcommand{\sO}{{\mfs O}}

\def\i{i}

\begin{document}

\title{Radiation from quantum weakly dynamical horizons in LQG}

\date{\today}

\author{Daniele Pranzetti$^1$}

\thanks{pranzetti@aei.mpg.de}

\affiliation{$^1$Max Planck Institute for Gravitational Physics (AEI), 
Am M\"uhlenberg 1, D-14476 Golm, Germany.}


\begin{abstract}
Using the recent thermodynamical study of isolated horizons by Ghosh and Perez, we provide a statistical mechanical analysis of isolated horizons near equilibrium in the grand canonical ensemble. By matching the description of the dynamical phase in terms of weakly dynamical horizons with this local statistical framework, we introduce a notion of temperature in terms of the local surface gravity. This provides further support to the recovering of the semiclassical area law just by means of thermodynamical considerations.
Moreover, it allows us to study the radiation process generated by the LQG dynamics near the horizon, providing a quantum gravity description of the horizon evaporation. For large black holes, the spectrum we derive presents a discrete structure which could be potentially observable and might be preserved even after the inclusion of all the relevant transition lines.

\end{abstract}


\maketitle



{\bf Introduction.} In order to fully understand and explain the semiclassical result of Hawking's calculation \cite{Hawking} regarding the emission of thermal radiation from a black hole, it is strongly believed that a theory of quantum gravity is necessary in order to take into account the quantum structure of the horizon geometry at the Planck scale. 

The issue of the dependence of Hawking thermal spectrum derivation on the behavior of the fields at arbitrarily high frequencies has been analyzed, for instance, by Jacobson \cite{Jacobson} and Unruh \cite{Unruh}, with the first one speculating that, due to the presence of a short distance cutoff, the spectrum may undergo modifications. 

The possibility to probe the Planck scale structure of black holes with observations at much bigger wavelength has been conjectured by Bekenstein and Mukhanov \cite{Beke}. Developing earlier proposal of theirs regarding black holes as discrete quantum systems, the authors start from the assumption that the area of the black hole be quantized and, due to the relation between black hole surface area and its mass, they study the emission amplitude related to the jump from one quantized value of the mass to a lower one, in analogy with atoms behavior. Due to the discreteness of the area eigenvalues, the full emission spectrum is given by a set of spectral lines at frequencies multiple of a fundamental one of order of the black hole surface gravity, whose envelope reproduces Hawking's thermal spectrum. This would imply a drastic departure from Hawking's semiclassical result, as emphasized by Smolin \cite{Smolin}.

However, Rovelli  et al. have shown \cite{Rovelli} that, replacing the naive ansatz of \cite{Beke} for the area spectrum with the one computed from LQG, the spacing between the energy levels decreases exponentially with the black hole mass and, henceforth, for a macroscopical black hole, the spectral lines are very dense in frequency, recovering in this way a continuos spectrum (see, however, \cite{Ansari} for a more detailed analysis based on the LQG area spectrum and \cite{Barrau} for a numerical investigation applied to microscopical black holes).


The `atomic' picture of a quantum black hole adopted in \cite{Beke} was further exploited by  
Krasnov \cite{Krasnov} within the LQG framework, soon after the first exciting ideas \cite{SRK} concerning the derivation of black hole entropy from the d.o.f. of the horizon quantum geometry started to circulate. Those ideas led to the introduction of a local definition of black hole suitable to a quantum gravity treatment, from which a microscopic origin of black hole entropy was derived. More precisely, through the quasilocal notion of {\it Isolated Horizons} (IH) \cite{IH}, able to capture the main physical properties of black holes event horizons, the LQG techniques were applied to space-time regions containing such an horizon, leading to a model \cite{ABCK, SU(2)} where the quantum black hole d.o.f. are described by a Chern-Simons theory on a punctured two-sphere. The punctures represent the quantum excitations of the gravitational field, as described by the 
LQG kinematics. Counting the dimension of the Hilbert space of such a theory living on the horizon leads to an entropy in agreement with the Bekestein-Hawking result, once the large area limit is taken. 

Krasnov's original idea was to use this quantum mechanical description of black hole states to study the emission and adsorption spectrum of quantum black holes. By analogy with the gas of atoms and the use of Fermi's golden rule, he analyzed the intensity of emission lines corresponding to a single puncture transition from spin $j$ to $j'$. The spectrum he finds shows a thermal character, even though it presents a discrete structure. However, in his analysis no dynamical process responsible for the puncture transition is taken into account. In particular, in the derivation of the spectrum he considers a single ÔatomÕ transition approximation; this leads to the awkward feature of removing the line $1/2 \rightarrow 0$, which, for macroscopic black holes, would represent the most likely transition.
Moreover, the statistical mechanical framework of \cite{Krasnov} lacks a clear connection with the usual energy canonical ensemble, which is the fundamental ingredient for the derivation of all the thermodynamic functions (see \cite{Barbero-Villa} for a recent analysis of the area canonical ensemble). 

In this letter, we want to investigate further and more in detail the analogy between a quantum horizon with its punctures and a gas of particles by introducing the main ingredients for a grand canonical ensemble analysis\footnote{See \cite{Major} for a previous grand canonical ensemble treatment of a gravitational system with a spatial boundary, where the Hamiltonian used in the definition of the partition function is the quasilocal energy associated with the boundary.}. The basic idea is to regard the bulk and the horizon as forming together an isolated system. The two subsystems are considered separately in thermal equilibrium; then, at some point, a weakly dynamical phase takes place and they interact with each other. This local interaction allows for the possibility of exchange of energy and particles between the two. After such a change of thermodynamic state has taken place, the two subsystems go back to a situation of thermal equilibrium. This picture will be made more precise in the following, where we will concentrate only on the spherically symmetric case. However, let us at this point clarify the framework we are working in: no background structure is introduced at any point, we will work in the quantum gravity regime; no matter is going to be coupled to gravity; the radiation spectrum we will derive is related entirely to emission of quanta of the gravitational field due to dynamical processes described by the LQG approach.

Now, in order to provide the tools for a thermodynamical study of the system described above, one has to introduce a notion of local energy for the horizon. 
As it is well known, in the context of IH there is no unique notion of quasilocal energy function. This is due to the fact that there can be radiation in space-time outside the horizon and, therefore, no unique time evolution vector field can be single out\footnote{Note that the same ambiguity is present also in the case of Killing horizons, since the freedom to rescale the Killing field normal to the horizon is still present. However, in the case of asymptotically flat space-times admitting global Killing fields, the normalization of the Killing field can be specified in terms of its behavior at infinity.}. 
Nevertheless, in \cite{AP, APE} a local first law for isolated horizons has been recently derived, whose uniqueness can be proven once a local physical input is introduced. Interestingly, the notion of energy associated to the isolated horizon for a preferred family of local observers $\sO$, at a proper distance $\ell$ from the horizon, turns out to be proportional to its area $A$, namely
\be\la{E}
E=\frac{\bar \kappa}{8\pi G} A,
\ee
where $\bar\kappa=1/\ell+o(\ell)$ represents the local surface gravity measured by the locally non-rotating stationary observer \cite{AP, APE}.
This result sets the basis for the study of the quantum statistical mechanics of IH started in \cite{AP}. 
\vskip0.3cm
{\bf   Entropy.} Let us now first concentrate on the derivation of the entropy of the gas of punctures (see \cite{entropy} for the original microcanonical derivation and \cite{Review} for a recent review). By working in the grand canonical ensemble---which represents the physically most suitable framework to describe the horizon+bulk system---, it can be shown how the Bekenstein-Hawking semiclassical entropy can be recovered only through thermodynamical considerations. 
Moreover, the description of the radiation process in the second part of the paper justifies the interpretation of the local notion of surface gravity introduced above as a temperature, which is a fundamental ingredient to recover the semiclassical area law (see below). In this sense, the result of the second part of the paper puts on more solid ground the recent derivation of \cite{AP}. This section simply presents a more detailed derivation of the IH entropy in the grand canonical ensemble already performed in \cite{AP}. The original part of the paper is contained in the next section.

The grand canonical partition function for the gas of punctures is given by
\be\la{gran}
\mfs Z(\beta)=\sum_{N=0}^\infty z^N Z(\beta,N)\,,
\ee
where $z=\exp{(\beta \mu)}$, $\mu$ being the chemical potential, $N$ the number of punctures and $Z(\beta,N)$ the canonical partition function given by\footnote{Here we are assuming the punctures statistics to be bosonic. We will comment more on this at the end of the paper.}
\be\la{can}
Z(\beta,N)=\sum_{\{s_j\}}\prod_j \frac{N!}{s_j!}(2j+1)^{s_j}e^{-\beta s_j E_j}.
\ee
In the previous expression, $\{s_j\}$ represents a given quantum configuration where $s_j$ punctures carry the spin-$j$, for all possible value of $j$, and such that $\sum_j s_j=N$, while $E_j$ is given by the scaled IH area spectrum through (\ref{E}), namely
\be\la{Ej}
\hat H |\{s_j\}\rangle=\sum_j s_j E_j|\{s_j\}\rangle= \frac{\bar \kappa\gamma \ell_p^2}{G}\sum_j s_j\sqrt{j(j+1)}|\{s_j\}\rangle.
\ee
Notice that, so far, $\beta$ can be regarded simply as the intensive parameter conjugate to the energy (\ref{E}) and its form is left unspecified. 
By means of the multinomial theorem, the grand canonical partition function (\ref{gran}) takes the form
\ba\la{gran2}
\mfs Z(\beta)&=&\sum_{N=0}^\infty\left(z\sum_j(2j+1)e^{-\beta  E_j}\right)^N\n\\
&=&\frac{1}{1-z\sum_j(2j+1)e^{-\beta  E_j}}.
\ea
Given the partition function, all the thermodynamical functions of the system can be computed from it. In particular, for the entropy we get
\be\la{Sgran}
S=-\beta^2\frac{\partial}{\partial \beta}(\frac{1}{\beta}\log \mfs Z)=\beta \bar E - \mu \beta \bar N+\log \mfs Z,
\ee
where
\be\la{bar}
\bar E = -\frac{\partial}{\partial \beta} \log \mfs Z~~~{\rm and}~~~\bar N= z \frac{\partial}{\partial z} \log \mfs Z 
\ee
are respectively the mean energy and the mean number of punctures of the gas. If we compute the expectation value for the occupation numbers $s_j$, namely
\be\la{Occup}
\bar s_j=-\frac{\partial}{\partial (\beta E_j)}\log \mfs Z= \frac{z(2j+1)e^{-\beta  E_j}}{1-z\sum_j(2j+1)e^{-\beta  E_j}},
\ee
it is immediate to see that equations (\ref{bar}) imply $\bar E=\sum_j \bar s_j E_j$ and $\bar N=\sum_j \bar s_j$. 
Therefore, the grand canonical entropy (\ref{Sgran}) can be written as
\be\la{Sgran2}
S=\frac{\beta\bar \kappa}{8\pi G}A - \mu \beta \bar N+\log \mfs Z,
\ee
where the first term in the r.h.s is the microcanonical entropy computed from the dominant configuration of the occupation numbers $\{s_j\}$ \cite{AP}. 
The expression (\ref{Sgran2}) can be seen as the horizon entropy at a generic temperature $\beta$ for the preferred local observer $\sO$ hovering outside the horizon at proper distance $\ell$, as resulting from the (physical) local version of the IH first law $d E= \bar \kappa/(8\pi G) d A$, which implies the local notion of energy (\ref{E}) \cite{APE}.

In order to recover the Bekenstein-Hawking result from the grand canonical entropy (\ref{Sgran2}), we need to look at its form in the semiclassical regime. First, we take the large $N$ limit; secondly, we need to introduce a physical input proper of the  black hole geometry outside, but near, the IH. 

When the number of punctures is big, the second of (\ref{bar}) gives 
\be\la{LarN}
1\approx\bar N/(\bar N+1)=z\sum_j(2j+1)e^{-\beta  E_j},
\ee
from which
\be\la{mu}
\mu\approx-1/\beta \log {(\sum_j(2j+1)e^{-\beta  E_j})}.
\ee
Notice that, by means of the expression (\ref{Ej}) for the energy levels, the sign and the vanishing of the chemical potential above become a function of $\gamma$; this may have important physical implications for different sectors of the value of $\gamma$, which, however, won't be investigated here. As pointed out in \cite{AP}, the chemical potential does not need to vanish in order to recover the semiclassical result, due to the quantum nature of the hair correction $\bar N\mu$. 
For the gas of photons description of black-body radiation, the vanishing of the chemical potential is consistent with the requirement that the free energy of the system be a minimum, 
since the average total number of particles in the gas is indefinite. In our case, $\bar N$ is not fixed a priori either, but, due to the dynamical processes taking place near the horizon (as described below), a change in the number of puncture implies a change of the horizon energy and, therefore, $\mu$ can be different from zero. However, semiclassical consistency now implies a quantum modification of the first law, where a chemical potential term has to be introduced due to this hair term in the entropy.

The last missing ingredient, in order to recover Bekenstein-Hawking area law \cite{Hawking, beke}, is now the identification of the radiating temperature $\beta$ with the inverse of the horizon surface gravity. Let us recall that this identification made sense only after Hawking's surprising result that the black hole could radiate; before that, it only had the status of a resemblance or a coincidence suggested by the equations form, since the black hole temperature was expected to be zero. 

In the second part of the paper, we will argue that quantum geometrodynamical considerations could generate a radiation process from the quantum horizon, for which the local surface gravity $\bar \kappa$ can be properly interpreted as a temperature. Despite being of a completely different nature (respect to the Hawking effect) and not specifying the proportionality coefficient, this process allows us to introduce a notion of temperature for the quantum IH. One can now fix this relationship between temperature and surface gravity using the semi-local (and semi-classical) argument of \cite{AP}. Namely, if we assume a stationary near-horizon geometry and we use the Unruh temperature $\beta \bar \kappa=2\pi/\hbar$ for our local accelerated observer $\sO$, the entropy expression (\ref{Sgran2}) gives exactly $S=A/(4\ell_p^2)$, at the leading order.
\vskip0.3cm
{\bf Radiation.} We now switch our attention to the radiation process. The topology of the null surface $\Delta$ representing the IH is assumed to be $S^2 \times \mathbb{R}$. This means that, at `a given time', the horizon surface is given by the two-sphere at the intersection between $\Delta$ and a Cauchy surface representing space. The quantum space geometry is described by the polymer-like excitations of the gravitational field encoded in the spin networks states, which span the kinematical Hilbert space of LQG. Some edges of those states can now pierce through the horizon surface, providing local quantum d.o.f. accounting for the IH entropy, as shown above. Dynamics in the bulk is implemented by the Hamiltonian constraint $\hat H[N]=0$. This reflects the fact that, in a diffeos invariant theory, the canonical Hamiltonian generates evolution in the parameter of the action which is unphysical, i.e. the evolution is pure gauge. This freedom in choosing the (unphysical) evolution parameter is reflected in the possibility to rescale the time evolution vector field by the lapse $N$. Then, in the quantum theory, the projection operator into the kernel of the Hamiltonian constraint is constructed by integrating over all the possible value of the lapse $N$ (\cite{Teitelboim, Projector}).


Despite the lack of a completely well defined quantization prescription for the Hamiltonian operator,
some general properties are satisfied by all the several proposals present in the literature. 
 Namely, $\hat H$ acts locally at the vertices of the spin networks states and its action changes the spin associated to some of the edges attached to the given vertex. In order to be more quantitative, in the following
we will consider Thiemann's proposal \cite{QSD}.
Therefore, if we concentrate on the Euclidean constraint, for simplicity, its action on a 3-valent node having two edges piercing the horizon can be graphically represented as
\be
\begin{array}{c}  \includegraphics[width=5.5cm,angle=360]{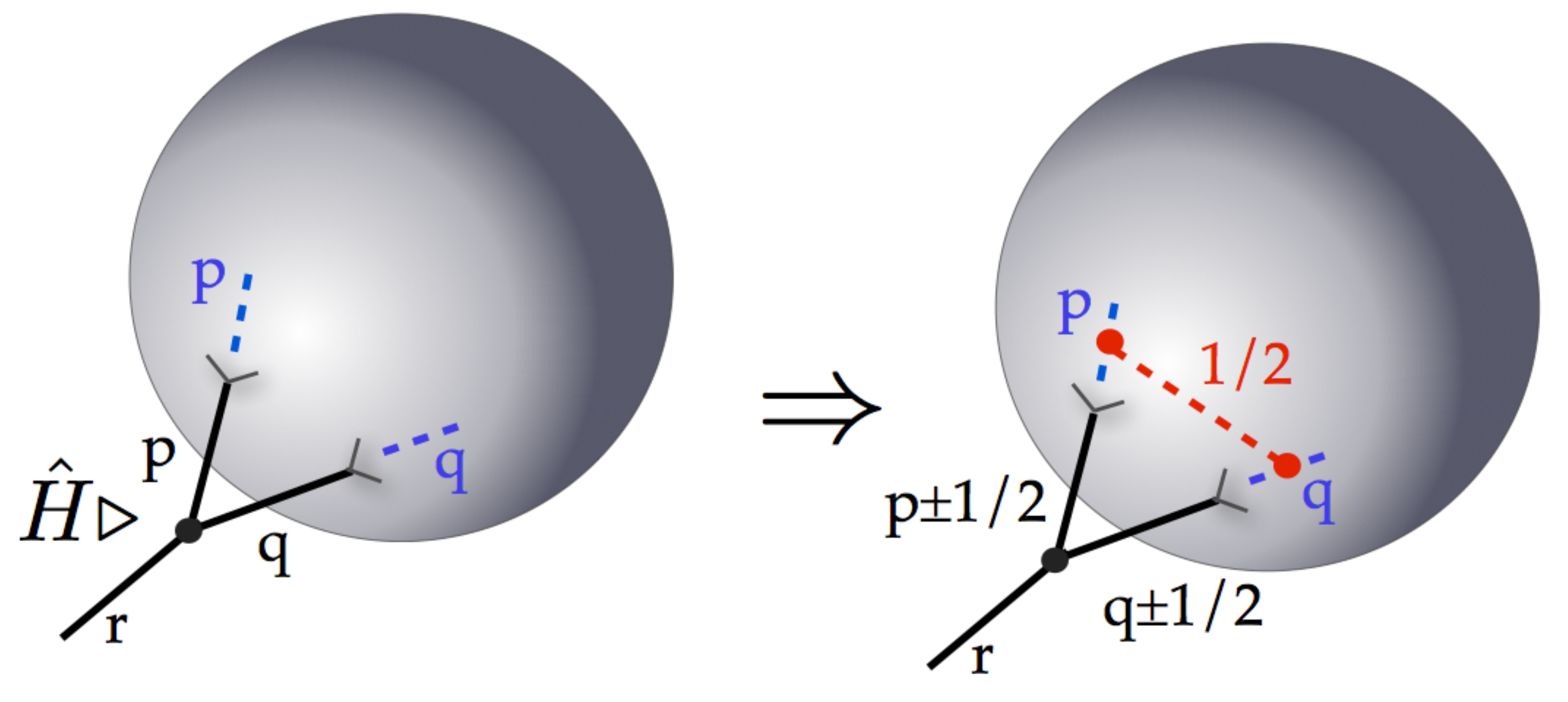}\end{array}~~~~~~~~~~~\begin{array}{c}  \includegraphics[width=1.7cm,angle=360]{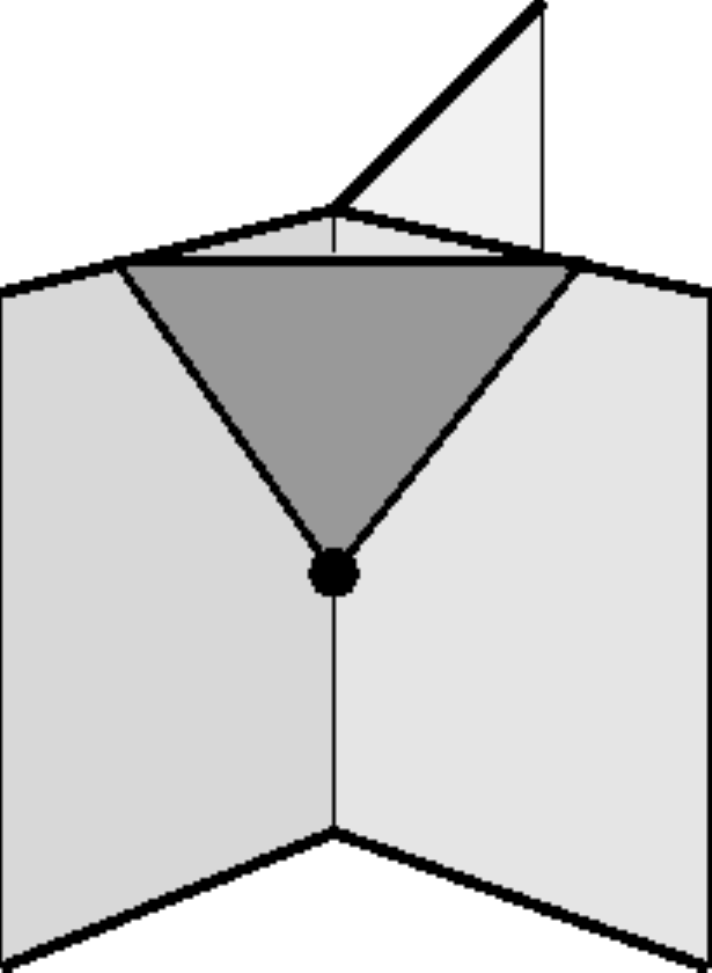}\end{array}\!,\la{H}
\ee
where the holonomies entering the regularization of $\hat H$ are taken in the fundamental representation (the fundamental vertex on the right provides the covariant representation of the Hamiltonian operator action). The action depicted above can be interpreted as an absorption/emission process of quanta of the gravitational field by the quantum horizon.

In order to study the radiation process, we need to `perturbate' the equilibrium states represented by the IH configurations by turning on some weakly dynamical effects near the horizon, till another equilibrium configuration state is reached again. In this way,  a physical process can causes the transition from one equilibrium state to a nearby one. The horizon goes through a non-equilibrium phase resulting in a small change of its state. For a large black hole, this process is expected to be quite slow. 

In order to describe this evolution process, we need first to deparametrize the system near the horizon, i.e. we need to locally single out one of the partial observables to play the role of time. In this way, we can use the Hamiltonian operator in the bulk near the horizon to define {\it evolution} of the boundary quantum states with respect to this observable.

\begin{figure}[h]\vspace{0cm}
\centerline{ \(\begin{array}{c}
\vspace{-0.3cm}\hspace{1.3cm}\includegraphics[height=3.3cm]{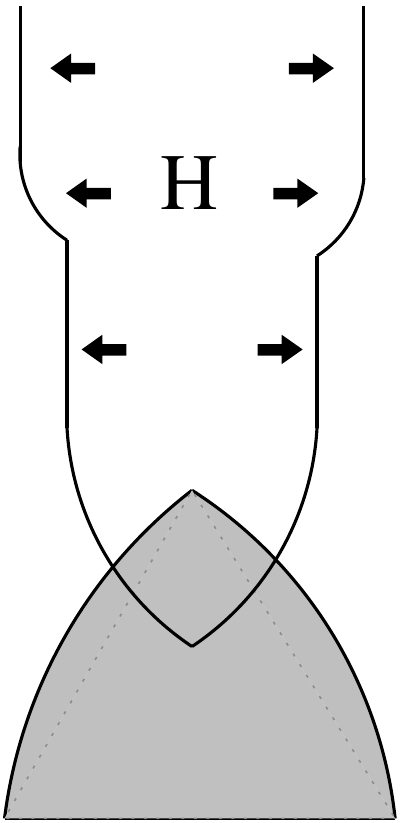}
\end{array}\!\!\!\!\!\!\!\!\!\!\!\! \!\!\!\!\!\!\!\!\!\!\!\!  \hspace{0.05cm}\begin{array}{c}\Delta_2  \\\\ \\    \vspace{1.5cm} \Delta_1
\end{array}
\hspace{1.0cm}\begin{array}{c}~~~~~~~~~~~~~~~~~~~~~~~~~~~~~~~~~~~~~~~~~~~~~~~~~\\
\includegraphics[height=2.1cm]{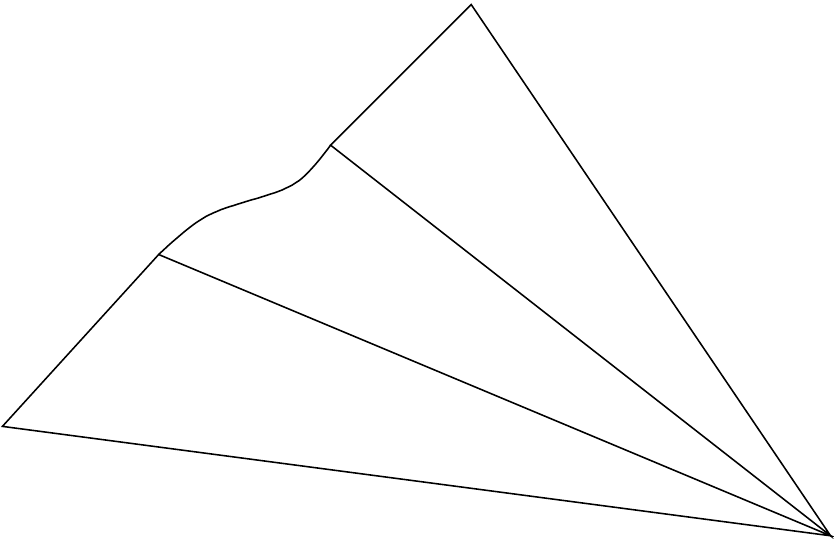}
\end{array} \hspace{0.0cm} \begin{array}{c}\!\!\!\!\!\!\!\!\!\!\!\!\!\!\!\!\!\!\!\!\!\!\!\!\!\!\!\!\!\!\!\!\!\!\!\!\!\!\!\!\!\!\!\!\!\!\!\!\!\!\!\!\!\!\!\!\!\!\!\!\!\!\!\!\!\!\!\!\!\!\!\!\!\!\!\!\!\!\! \!\!\!\!\!\!\!\!\!\!\!\!\!\!\!\! \!\!\!\!\!\!\!\!\!\!\!\!\!\!\!\!\!\!\!\!\!\!\!\!\!\! \!\!\!\!\!\!\!\! \!\!\!\!\!\!\!\!\!\!\!\! \!\!\!\!\!\!\!\! \!\!\!\!\!\!\!\!\!\!\!\! \!\!\!\!\!\!\!\! \!\!\!\!\Delta_2  \!\!\!\!\!\!\!\!\!\!\! \!\!\!\!\!\!\!\! \! \!\!\!\!\!\!\!\!\!\!\!\!\!\!\!\!\!\!\!\!\!\!\!\!\!\!\!\!\!\! \\\!\!\!\!\!\!\!\!\!\!\!\!\!\!\!\!\!\!\!\!\!\!\!\!\!\!\!\!\!\!\!\!\!\!\!\!\!\!\!\!\!\!\!\!\!\!\!\!\!\!\!\!\!\!\!\!\!\!\!\!\!\!\!\!\!\!\!\!\!\!\!\!\!\!\!\!\!\!\!\!\!\!\!\!\!\!\!\!\!\!\!\!\!\!\!\!\!\!\!\! \!\!\!\!\!\!\!\!\!\!\!\!\!\!\!\!\!\!\!\! \!\!\!\!\!\!\!\!\!\!\!\!\!\!\!\! \!\!\!\!\!\!\!\!\!\!\!\!\!\!\!H\!\!\!\!\!\!\!\!\!\\ \\ \!\!\!\!\!\!\!\!\!\!\!\!\! \!\!\!\!\!\!\!\! \!\!\!\!\!\!\!\!  \!\!\!\!\!\!\!\!\!\!\!\!\!\!\!\!\!\!\!\!\!\!\!\!\!\!\!\!\! \! \!\!\!\!\!\!\!\!\!\!\!\!\!\!\!\! \!\!\!\!\!\!\!\!\!\!\!\!\!\!\!\! \!\!\!\!\!\!\!\!\!\!\!\!\!\!\!\! \!\!\!\!\!\!\!\!\!\!\!\!\!\!\!\! \!\!\!\!\!\!\!\!\!\!\!\!\!\!\!\! \!\!\!\!\!\!\!\!\!\!\!\!\!\!\!\! \!\!\!\!\!\!\!\!\!\!\!\!\!\!\!\! \Delta_1
\end{array}\) }\vspace{-0.05cm} \caption{After the collapse of a spherical star, the IH $\Delta_1$ is created. Later, in the dynamical phase $H$, flux of gravitational and matter energy goes through the horizon, before a new equilibrium state $\Delta_2$ is reached again.}\vspace{-0.1cm}
\label{IH}
\end{figure}

Let us now show how the contact between the {\it Dynamical Horizons} framework developed in \cite{DH1, DH2} and the thermodynamical description of \cite{AP, APE} allow us to single out such a time evolution vector field. Within a framework advocated by Ashtekar, the idea is to describe the evaporation process by interpreting this intermediate phase of horizon state evolution as an extension of the quantum horizon geometry from isolated to dynamical
horizons, where a proper quantum notion of gravitational energy---induced by the quantum theory of geometry in the bulk---replaces the classical one introduced in \cite{DH2, Booth}.


In \cite{DH2}, a dynamical process version of the first law was derived from an {\it area balance law} relating the change in the area of the dynamical horizon to the flux of matter and gravitational energy. In the absence of matter, this could be written as
\be\la{balance}
\left(\frac{r_2}{2 G}-\frac{r_1}{2 G}\right)=\frac{1}{16 \pi G}\int_H N_r H_{\va B} d^3V,
\ee
where $H$ is the portion of dynamical horizon bounded by the 2-sphere leaves $S_1, S_2$ of radius $r_1, r_2$ and $H_{\va B}$ is the bulk term in the Hamiltonian, which can be written in terms of pure geometrical quantities\footnote{Notice that the l.h.s. of (\ref{balance}) is obtained from a boundary term in the Hamiltonian, as shown already in \cite{ABF}.}. The infinitesimal version of (\ref{balance}) gives a first law for dynamical horizons, namely
\be\la{first}
\frac{\bar \kappa_r}{8\pi G} dA= dE_r,
\ee
where $ \kappa_r=(2r)^{-1}$ is an {\it effective} surface gravity.

First of all, let us notice that, given the spherical symmetry of the horizons studied here, the classical notion of gravitational energy defined by the r.h.s. of (\ref{balance}) would give a vanishing flux across the dynamical horizon. Of course, this doesn't mean that also at the quantum level there cannot be a flux of energy across the horizon. Hawking radiation taught us exactly the opposite. 

In \cite{DH2} it is shown that, for a dynamical horizon, the first law (\ref{first}) can be generalized to $\bar \kappa_f/(8\pi G) dA=dE_f$, where $\bar \kappa_f=(df/dr) \bar \kappa_r$, with $f(r)$ an arbitrary function of the radius $r$ encoding the freedom to rescale the vector field $\xi_f$---with respect to which the notion of energy is associated\footnote{In the case of IH, this corresponds to the freedom to rescale the null normal vector field $\chi^a$, and therefore the surface gravity, by a positive constant $c$.}---, i.e. $\xi^a_f=N_f \chi^a=(df/dr) \xi^a_r$, where $N_f$ is a permissible lapse associated with the radial function $f$. The function $f(r)$ encodes the dynamical nature of the previous version of the first law, since $r$ plays the role of time along the dynamical horizon; but it also represents an ambiguity in the description of the horizon dynamics.

The passage from a weakly dynamical horizon $H$ to a weakly isolated horizon $\Delta$ is a well defined procedure described in \cite{DH2}. The only ingredient missing in our construction is the matching of the surface gravity entering (\ref{first}) and the one appearing in the local notion of energy (\ref{E}). This last step can now be carried out by means of the rescaling freedom described above, using the local universal first law of \cite{APE} to solve this ambiguity.

For our preferred family of observers located right outside the horizon, a natural choice of the field $\xi_f$ would be given by $f=\ell$, with $\ell$ being the proper distance from the horizon\footnote{In the generalization of the area balance law introduced in \cite{DH2}, the function $f$ is assumed to be constant in each 2-sphere cross section of the dynamical horizon $H$. In this way, the permissible vector field $\xi_f$ is such that the flux of gravitational energy, associated to it, across a portion $\Delta H$ on the horizon relates the two constant values the function $f$ assumes on the fixed cross-sections bounding $\Delta H$. Since a change in the horizon radius implies a change in the proper distance of the local observer from one cross-section to another, our choice $f=\ell$ is compatible with the generalization of \cite{DH2}.}. If we use the fact that, in standard coordinates for the Schwarzschild metric, for an observer at $r=2M+\epsilon$, the proper distance from the horizon is $\ell=2\sqrt{2M\epsilon}$ and set $f(r)=\ell$, we get an effective surface gravity matching exactly the local surface gravity entering the energy expression (\ref{E}), i.e. $\bar \kappa_f=1/\ell$ \footnote{The same result can be obtained without referring to any global structure by using the local extension of the null normals $\chi^a$ to the vicinity of the horizon, as constructed in \cite{APE} and used to define the preferred local observer $\sO$.}. Therefore, the local first law for black hole thermodynamics derived in \cite{AP, APE} is compatible with the framework of \cite{DH2} if we single out, among the permissible time vector fields for the local observer $\sO$, the one defined by $t^a=N_\ell \chi^a$ .

Moreover, if we interpret $\epsilon$ as a parameter controlling the slow evolution of the horizon area, i.e. $\sL_\chi r_{\va H}\sim \epsilon$, the previous expression for the surface gravity $\bar \kappa_f$ matches the requirements of \cite{Booth} for the definition of slowly evolving horizons. In this case, the local surface gravity entering the relation (\ref{E}) would be approximately constant. Henceforth, one can regard the horizon as making continuous transitions from one equilibrium state to another (see FIG. \ref{IH}) and the geometrical surface gravity $\bar \kappa$ can be interpreted as a (slowly varying) parameter controlling this process. In this sense, it seems very natural to interpret $\bar \kappa$ as the temperature of the radiation emitted during this transition phase. This strengthens the connection between thermodynamics and our statistical analysis and clarifies  further the entropy derivation performed in \cite{AP}. 

At the quantum level, the local notion of energy (\ref{E}) acquires a definition in terms of the LQG area operator: this is the fundamental ingredient for the construction of the quantum statistical mechanical ensemble above. Since the Hamiltonian operator acts on spin network states and modifies them, we can now use the local deparametrization of the system described above to interpret the bulk quantum dynamics near the horizon as a generator of boundary states evolution. In other words, we can now make sense of the r.h.s. of (\ref{first}) at the quantum level by means of the LQG dynamics. The consequent jump to a different horizon area operator eigenstate would thus result in an horizon energy variation. In this way, quantum dynamics would allow for an energy flux across the horizon, providing a quantum notion of weakly dynamical horizon.

More precisely,
the action of the Hamiltonian operator on nodes close to the horizon (see (\ref{H})) can be used, in the formalism of \cite{Projector}, to define a physical scalar product to interpret as transition amplitudes between one boundary state to another, in the physical time evolution parameter defined by $N_\ell$. In general, given a quantum boundary state $|\{s_j\}\rangle$ on a given 2-sphere cross-section, the action of the Hamiltonian operator on the vertices near the horizon will evolve it in a state $|\{s_{j'}\}\rangle$ through the change of spin of some of the punctures in the initial state $|\{s_j\}\rangle$.
The probability of the transition between an initial state $|\{s_j\}\rangle$ and a final state $|\{s_{j'}\}\rangle$ differing only at a given vertex $v$, in an infinitesimal interval of time, is given by $P_{jj'}= |H_{jj'}|^2$, where
\ba\la{scalar}
H_{jj'}&\equiv&\langle \{s_{j'}\}|\{s_j\}\rangle_{phys}=
\langle \{s_{j'}\}|\int dN O_\ell e^{-\i \hat H[N]}|\{s_j\}\rangle\n\\
&=&-\i N_\ell \langle \{s_{j'}\}|  \hat H_v|\{s_j\}\rangle+o(N_\ell^2).
\ea
Let us explain the last passage in the previous equation. In order to define a diff-invariant physical scalar product, the action of $\hat H$ is defined on states of $\sH_{aux}$, hence picking up a dependence on the regulator (i.e. the position of the link added); however, this dependence finally drops out in the integration over $N$. More precisely, one can insert inside $\langle\cdot|\cdot\rangle_{phys}$ an integration over the diffeomorphisms and this confines the remaining diff-dependent quantities in the arguments of the functions $N(x)$; the diff-invariant nature of the $DN$ integral then eliminates any residual dependence. In the case of an internal boundary though, the situation is more subtle. In particular, IH boundary condition are such that only diffeos tangent to the horizon are to be regarded as gauge transformations in the Hamiltonian framework \cite{IH}; these diffeos are compatible with the `gauge fixing' performed in the evaluation of (\ref{scalar}). This gauge-fixing, which encodes the local deparametrization of the system due to the choice of a preferred (`area') time variable described above, could be performed by inserting a delta function in the measure of the $DN$ integral which singles out the constant value $N_\ell$ for the scalar function $N(x)$ at the vertices near the horizon. This procedure could be interpreted as the insertion of a local observable $O_\ell=\delta(N,N_\ell(v))$ in the path integral representing our physical time-evolution variable.

Let us now concentrate on the description of the radiation process. In defining $H_{jj'}$, we will use the matrix elements of the Hamiltonian operator obtained in \cite{Matrix}, where only the Euclidean term, $\hat H^{\va E}$, of Thiemann's version \cite{QSD} of the constraint is considered. Eventual modifications of the radiation spectrum induced by the inclusion of the `Lorentzian' term are left for future study.



At this point, an important observation concerns the local nature of the dynamical process near the boundary in relation with the action of the Hamiltonian operator. 
The terms generated by the action of $\hat H^{\va E}$ on a generic 3-vertex near the horizon are of two  different kinds in our context: two of them create new punctures piercing through the horizon and one (that shown in (\ref{H})) does not. In the first case, the horizon area can only increase or at most remain constant (no radiation). The inclusion of those terms creating new punctures though would spoil the local characterization of the interaction between the bulk and the horizon, by introducing an arbitrary dependence of the transition amplitudes on the spin of the bulk links adjacent to the interaction vertex. Moreover, since the new puncture created by these terms has to be co-planar to the two it is attached to (one piercing the horizon and one in the bulk), the entropy calculation for the new equilibrium configuration may lead to divergences, due to the fact that one can no longer trace over the bulk d.o.f. without affecting the horizon state and the moding out of the diffeos tangent to the horizon becomes non trivial. 

For these reasons, we only allow the action of $\hat H^{\va E}$ that doesn't create new punctures. Notice that this term corresponds to the absorption inside the horizon of a quantum of the gravitational field (the new link created by $\hat H^{\va E}$) associated to an emission process, due to the jump of the horizon state to a lower area eigenvalue. This dynamics is reminiscent of the heuristic picture of Hawking radiation where one anti-particle is absorbed by the black hole while one particle escapes to infinity. In this sense, the original idea \cite{Hawking} that Hawking had in mind, to describe the black hole evaporation, is realized in our context.




Given these considerations, the $H_{jj'}$ in (\ref{scalar}) corresponds to the transition amplitude where two of the punctures piercing the horizon have jumped to a different energy level. Since we want to study the emission process, we will now take into account only the jumps to a lower eigenvalue of the area (energy) operator\footnote{This breaking of unitariety can be justified by the fact that we are looking only at a part of the system (the near-horizon dynamics), independently of what happens in the bulk. We thank C. Rovelli for pointing this out.}. 

For the case of a macroscopic black hole considered here, only the first $SU(2)$ Irreps are relevant for the horizon punctures, henceforth, we will consider only the cases where $q=j$ and $p=j,j+1/2, j+1, j+3/2$, i.e. $r=1/2, 1, 3/2$ ($r$ being the spin of the third link at the 3-vertex acted upon not piercing the horizon). Higher values of $r$ would not be relevant for the spectrum of a large black hole; this is a feature of the action of $\hat H$ when no new punctures are created, as observed above.
We now choose the holonomies in $\hat H^{\va E}$ to be in the fundamental representation, implying that $q$ can only jump to $q-1/2$ while $p$ to $p\pm 1/2$. 

Another source of ambiguity is represented by having two possible orderings of $\hat H^{\va E}$, as defined by Thiemann. More precisely, the volume term inside $\hat H^{\va E}$ can act before or after the creation of the new link. Here we will choose the ordering in which the volume operator acts before the punctures have jumped in order to have a non vanishing probability for the transitions involving the jump $1/2 \rightarrow 0$ for one of the punctures.
This will be our choice for the regularization of the Hamiltonian operator (we will comment below on the ambiguity of choosing different representations).

Notice that, with this `two atoms' approximation motivated by the action of $\hat H^{\va E}$, the jump $j\rightarrow j-1/2$ involves always an even number of punctures; in this way, the gauge symmetry of the IH is preserved, as an equilibrium state is reached again, and the unpleasant feature of removing the $1/2 \rightarrow 0$ transition from the spectrum is not present anymore.


In the expansion (\ref{scalar}) of the transition amplitude, we have kept only the first order term in $N_\ell$, corresponding to a single action of the Hamiltonian operator at $v$;
this encodes at the quantum level the weak time dependence of the horizon in this intermediate phase between two equilibrium configurations. The contemporary action of $\hat H^{\va E}$ on more than one node would produce a bigger number of lines in the spectrum, allowing for the possibility that the discrete structure of the spectrum derived below get smoothen, resulting in an effective continuos spectrum. However, there is the possibility that these lines would be significantly suppressed by the higher powers of the transition amplitude $|N_\ell \langle  \hat H^{\va E}_v\rangle|$, preserving in this way the discrete structure. Further investigation of this feature is crucial for a possible observation of Planck's scale effects. 


The $\hat H^{\va E}$ matrix elements $A(2q,\bar\epsilon; 2q, \tilde \epsilon;2r)$ corresponding to the jumps $2p\rightarrow 2p+\bar\epsilon, 2q\rightarrow 2q+\tilde \epsilon$, with $\bar\epsilon,\tilde \epsilon=\pm1$, are explicitly computed in \cite{Matrix}.

We now have all the ingredients to derive the spectrum form, for our local observer $\sO$, of the radiation emitted by the quantum horizon in the process $|\{s_j\}\rangle\rightarrow |\{s_{j'}\}\rangle$. Using the analogy with the gas of particles, the eigenvalues (of the operator corresponding to (\ref{E})) $E_j$ represent the energy levels of the system, which are occupied by $\bar s_j$ quanta of the gravitational field. In our picture then, the punctures play the role of the photons in the Bose-Einstein approach to the investigation of {\it black-body radiation} and $E_j$ is the analog of a photon energy level $\hbar \omega_p$. 
According to the emission process described above, the energy of the quantum of radiation emitted after the action of $\hat H^{\va E}$ on two punctures with spins $p,q$ is given by $\Delta_{pq}^\pm=E_p-E_{p\pm\frac{1}{2}} +E_{q}-E_{q-\frac{1}{2}}$. 
Therefore, the intensity of the spectrum lines is given by
\vspace{-0.2cm}
\be\hspace{1.2cm}
I_{jr}=\bar s_p \bar s_q |N_\ell A(2p,\pm1;2q,-1,r)|^2\Delta_{pq}^\pm,
\ee
where, with the convention introduced above, $q=j, j\leq p\leq j+r$ and $j,r=1/2,1,3/2$; the occupation numbers are given by eq. (\ref{Occup}).



\begin{figure}[h]
\centerline{ \(
\begin{array}{c}
\includegraphics[height=5.cm]{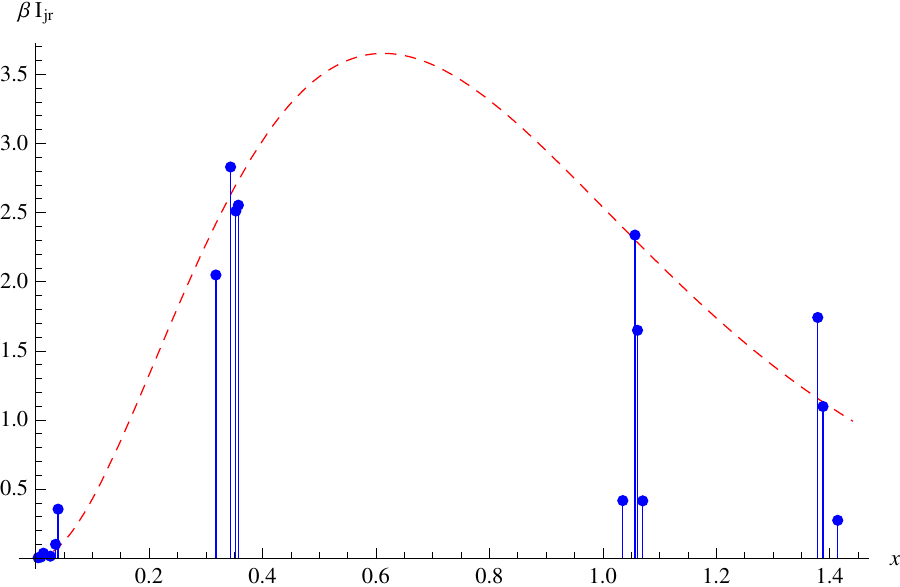}
\end{array}\) }
 \caption{Emission lines with their intensities and the thermal envelope. The normalization factor $N_\ell (\bar N+1)$ of the intensity is left unspecified since we are interested only in the form of the spectrum.}\vspace{-0.2cm}
\label{Spectrum}
\end{figure}

The total energy emitted in the process is given by $I=\sum_{j,r} I_{jr}$.
In FIG. \ref{Spectrum} we plot the energy distribution $I_{jr}$ as a function of $\beta$ times the energy of the quanta emitted, i.e. $x\equiv \beta\Delta_{pq}^\pm$ \footnote{Here we have used the interpretation of the surface gravity as the radiation temperature discussed above and the fact that, in the large $\bar N$ limit, $\beta \bar \kappa\gamma \ell_p^2/G\sim o(1)$.}. 
The plot shows a discrete structure formed by four different sets of spectral lines. The first two groups of lines correspond to the case $\bar \epsilon=+1$, i.e. when $p\rightarrow p+1/2$, while the last two to the case $\bar \epsilon=-1$, i.e. $p\rightarrow p-1/2$. 

The envelope of the emission lines reproduces a thermal behavior. 
However, since the frequency gap between the different sets of lines is of the order $\Delta \omega\sim  \bar\kappa$, a departure from a continuos spectrum for macroscopic black holes could be potentially observable.
\vskip0.5cm
{\bf Conclusions \& comments.}
We have used the statistical mechanical framework introduced in \cite{AP, APE} to study the properties of IH at and near the equilibrium in the grand canonical ensemble. This setting seems to be the most appropriate, given the possibility of the horizon to exchange energy with the bulk system, seen as a large reservoir, and the indeterminacy in the mean number of punctures. By matching the description of the intermediate dynamical phase in terms of weakly dynamical horizons \cite{DH2, Booth} with the local statistical description of IH in \cite{AP, APE}, we could introduce a notion of temperature regulating this exchange in terms of the local surface gravity $\bar \kappa_\ell$. Moreover, this  allowed us to single out a physical time parameter in terms of which describing the boundary states evolution.  In this way, our analysis provides further support to the main result of \cite{AP}, where 
the semiclassical area law for IH entropy is recovered just by means of thermodynamical considerations.

The kinematical Hilbert space of LQG provides a beautiful description of the quantum horizon d.o.f.. By means of the regularization and quantization prescription of the theory for the Hamiltonian constraint, we managed to define a quantum notion of gravitational energy flux across the horizon. This allowed us to study the radiation process generated by the Hamiltonian operator, providing a quantum gravity description of black hole evaporation. For large black holes, the discrete structure of the spectrum could be potentially observable. 

We conclude with a few comments. In analogy with the criticism of \cite{Rovelli} to the Bekenstein-Mukhanov effect, one could ask if the inclusion of other transitions between area eigenvalues would make the spectrum in FIG. \ref{Spectrum} effectively continuos.
According to the LQG dynamics, one way of adding more lines would be to go beyond the two atoms approximations and consider the simultaneous action of $\hat H$ on more than one vertex; as observed above, however, those lines could be significantly damped by the transition amplitude (\ref{scalar}), leaving the discrete structure unaltered\footnote{This would be the case, for instance, if $N_\ell \bar N\sim o(1)$.}. 

Another possibility would be to use the ambiguity in the holonomies representation present in the definition of $\hat H$. For large black holes, only the first few Irreps beyond the fundamental one would be relevant for the analysis. As it can be easily checked, due to the selection rules implemented by the Hamiltonian operator, the inclusion of the first lowest Irreps produces only a few more sets of lines whose frequencies, except for a couple, are completely shifted from (and with the same pattern of) the ones given by the fundamental representation (with a gap again of order $\sim \bar\kappa$) or exactly overlap to them. Therefore, even though the shape of the spectrum might change by including more transition lines, the LQG dynamics might   preserve its discrete structure, providing the possibility to observe a departure from the semiclassical scenario. However, further investigation in this direction is necessary and our derivation of the emission lines plotted in FIG. \ref{Spectrum} should be regarded as a first step towards a fully quantum dynamical derivation of the radiation spectrum.

While our attention here has been limited to Thiemann's proposal for the regularization of the Hamiltonian operator, the same analysis could be applied to different regularization schemes, such as \cite{Alesci}. The implications of alternative quantization of the Hamiltonian operator for the spectrum could eventually be used to discriminate between the different proposals and, hopefully, to reduce the number of ambiguities left in the definition of $\hat H$. 

Moreover, the near-horizon dynamics could be implemented within the spin foam formalism. More precisely, given the increasing evidence for the spin foam amplitude to implement the Hamiltonian constraint in the canonical
formalism \cite{SF}, one could use the vertex amplitude to compute transitions probabilities between boundary states containing the right part of (\ref{H}) as a subgraph. This could turn out to be a very useful application of the conjectured duality between the canonical and the covariant theory, with potential observational evidence. 
The analysis of these interesting issues is left for future study.

Despite the implementation of the Hamiltonian constraint to evolve the boundary quantum states, our analysis is still an effective one, since based on an interplay between classical general relativity and quantum geometry. It would be very interesting to apply the ideas developed here to the framework of \cite{Sahlmann}, where the quantum horizon d.o.f. are described within the full LQG formalism. This might provide further insights on the dynamical nature of the radiation process and could reveal unexpected connections between three and four dimensional quantum gravity. Moreover, the analysis of \cite{Sahlmann} seems to indicate that the punctures must obey some kind of anyonic statistics; investigating the consequences that such a different statistics would have on the radiation description could lead to important deviations from the bosonic assumption.  

Finally, on the relation with Hawking radiation. The two processes are quite different from each other. Hawking effect is `much less' dynamical \cite{Visser}; it simply requires a curved background and a scalar field propagating on it: the gravitational d.o.f. are frozen; the emission process is intrinsically related to the {\it definition} of particle possible due to this switching off of gravity. In our analysis, no background is ever picked up and no matter is coupled: gravity does all the job. It is the evolution of the quantum gravitational d.o.f. on the horizon, together with its beautiful interplay with classical general relativity and thermodynamics, which is responsible for the energy emission. Moreover, due to this dynamical framework, the whole description of the process is local (no notion of future/past null-infinity is involved). 

On the other hand, as Hawking radiation is a genuinely quantum effect, the emission process described here is also purely quantum. As noted above, the classical first law (\ref{balance}) derived in \cite{DH2, Booth} would give a vanishing flux for spherically symmetric horizon; it's the evolution---described by the LQG dynamics---of fluctuating quantum d.o.f. that generates radiation.

The analysis presented here could be extended to the inclusion of matter, whose d.o.f. are also expected to live on spin networks edges \cite{Matter}. Such an extension would provide a more appropriate set-up for a comparison with the Hawking effect. This is also left for future investigation. 


\vskip0.3cm
{\bf Acknowledgements.} I am very grateful to D. Oriti, A. Perez and L. Sindoni for clarifying discussions and important remarks concerning different aspects of this work.


\end{document}